\documentclass[12pt]{article}
\usepackage{amssymb}
\usepackage{amsmath}
\usepackage{multirow}
\usepackage{arydshln}

\usepackage[normalem]{ulem}

\numberwithin{equation}{section}
\newcommand{\be}{\begin{eqnarray}}
\newcommand{\ee}{\end{eqnarray}}
\newcommand{\non}{\nonumber}
\newcommand{\id}{\mathbb{I}}
\newcommand{\tr}{\mathop{\rm tr}\nolimits}

\usepackage[usenames,dvipsnames]{color}

\newcommand\blfootnote[1]{%
  \begingroup
  \renewcommand\thefootnote{}\footnote{#1}%
  \addtocounter{footnote}{-1}%
  \endgroup
}

\begin{document}

\begin{titlepage}
\strut\hfill UMTG--289
\vspace{.5in}
\begin{center}

\LARGE 
An inhomogeneous Lax representation\\
for the Hirota equation\\
\vspace{1in}
\large 
Davide Fioravanti \footnote{
Sezione INFN di Bologna, Dipartimento di Fisica e Astronomia, 
Universit\`a di Bologna, Via Irnerio 46, 40126 Bologna, Italy}
and Rafael I. Nepomechie \footnote{
Physics Department,
P.O. Box 248046, University of Miami, Coral Gables, FL 33124 USA}\\[0.8in]
\end{center}

\vspace{.5in}

\begin{abstract}
    Motivated by recent work on quantum integrable models without
    $U(1)$ symmetry, we show that the $sl(2)$ Hirota equation admits a Lax
    representation with inhomogeneous terms.  The compatibility of
    the auxiliary linear problem leads to a new consistent family of Hirota-like
    equations.
\end{abstract}

\blfootnote{e-mail addresses: {\tt davide.fioravanti@bo.infn.it, nepomechie@physics.miami.edu}}

\end{titlepage}

\setcounter{footnote}{0}

\section{Introduction and summary}
 
The Hirota equation (T system) is ubiquitous in the theory of quantum integrable
systems \cite{Hirota:1981, Kulish:1981bi, Kirillov:1987zz, Bazhanov:1987zu, Klumper:1992vt,
Kuniba:1993cn, Bazhanov:1994ft, Fioravanti:1995cq, Krichever:1996qd, Zabrodin:1998, Gromov:2008gj,
Gromov:2009tv}. For an $sl(2)$-invariant periodic quantum spin chain
(see Appendix \ref{sec:closedtransf} for details), the Hirota equation
takes the form \cite{Zabrodin:1998, Gromov:2008gj} \footnote{In general, 
the transfer matrix $T_{a,s}$ can have two subscripts, corresponding to the 
representation of the auxiliary space given by a rectangular Young tableau
with $s$ rows and $a$ columns. For simplicity, we focus here exclusively on the $sl(2)$ case, where $T$ has a single 
subscript $T_{s} = T_{1,s}$.}
\be
T^{+}_{k}\, T^{-}_{k} - T_{k+1}\, T_{k-1} = \phi^{[k]}\, 
\bar{\phi}^{[-k]}\,, \qquad T_{-1}=0\,,
\qquad k = 0, 1, 2, 
\ldots \,, 
\label{Hirota}
\ee
where $f^{\pm}= f(u \pm \frac{i}{2})$ and $f^{[\pm k]}= f(u \pm \frac{i k}{2})$
for any function $f(u)$. 
Here $T_{k}(u)$ is the
transfer matrix constructed with a spin-$k/2$ auxiliary space 
\cite{Kulish:1981bi, Kulish:1981gi}; in particular, $T_{1}(u)$ is the
fundamental transfer matrix.
These transfer matrices mutually commute $(\left[ T_{k}(u)\,, T_{j}(v) 
\right]=0)$, and obey the Hirota equation (\ref{Hirota}) with  
\be
\phi(u)=(u+\tfrac{i}{2})^{N}\,, \qquad 
\bar\phi(u)=(u-\tfrac{i}{2})^{N}\,.
\label{phi}
\ee 
The eigenvalues corresponding to simultaneous eigenvectors
of these transfer matrices (which we also denote by $T_{k}(u)$) 
evidently also obey the Hirota equation, and we henceforth regard $T_{k}(u)$ as a scalar function.

It is well known that the Hirota equation (\ref{Hirota}) admits a Lax representation
through the auxiliary linear problem (see \cite{Krichever:1996qd, 
Gromov:2008gj} and references therein) 
\be
T_{k+1}\, Q^{[k]} - T_{k}^{-}\, Q^{[k+2]} &=& \phi^{[k]}\, 
\bar{Q}^{[-k-2]} \,, \label{QI}\\
T_{k-1}\, \bar{Q}^{[-k-2]} - T_{k}^{-}\, \bar{Q}^{[-k]} &=& 
-\bar{\phi}^{[-k]}\, 
Q^{[k]} \, ,  \label{QII}
\ee
where the function $\bar{Q}(u)$ is defined by
$\bar{Q}(u)= \left(Q(u^*)\right)^{*}$. 
However, in order to reproduce the celebrated Baxter T-Q relation, we 
henceforth restrict our attention to the case that $Q$ is real analytic: 
$Q(u)^*= Q(u^*)$, which implies that
\be
Q(u)= \bar{Q}(u)
\label{QbarQ}
\ee
for any complex $u$.  Note that the Hirota equation (\ref{Hirota}) with $k=0$ can be satisfied by setting
\be
T_{0}=\phi^{-}\,, \qquad \bar\phi=\phi^{[-2]} \,.
\label{k0constraint}
\ee
It then follows from the first Lax equation (\ref{QI}) with $k=0$ (or 
alternatively from the second Lax equation (\ref{QII}) with $k=1$)
that
\be
T_{1}\, Q   = \bar{\phi}\, Q^{[2]} + \phi\, Q^{[-2]} \,,
\label{BaxTQ}
\ee
which is the important Baxter T-Q equation \cite{Baxter1982}.
Assuming that $Q(u)$ is a polynomial in $u$ of order $M$, i.e. 
$Q(u)=\prod_{j=1}^{M}(u-u_{j})$, the analyticity of $T_{1}$ together 
with the T-Q equation imply the Bethe equations for the zeros of 
$Q(u)$
\be
\bar{\phi}(u_{k})\, Q(u_{k}+i) + \phi(u_{k})\, Q(u_{k}-i) = 0\,,
\ee 
which can be rewritten in the more familiar form
\be
\left(\frac{u_{k}+\frac{i}{2}}{u_{k}-\frac{i}{2}}\right)^{N} = 
\prod_{j=1, j \ne k}^M
\frac{u_{k}-u_{j}+i}{u_{k}-u_{j}-i} \,.
\ee 
In principle, by solving the Bethe equations, one can obtain $Q$, and therefore 
(through (\ref{BaxTQ})) $T_{1}$.
Since (\ref{QI}) is linear, it can be solved for all the $T_{k}$ in terms of $Q$, 
and therefore it gives T-Q-like equations for the higher 
(fused) transfer matrices.

The conventional wisdom has been that (\ref{QI})-(\ref{QII}) is the
unique Lax representation for the Hirota equation.  However, it has
recently been shown that quantum integrable models without $U(1)$
symmetry (such as the open XXX spin-1/2 chain with non-diagonal
boundary terms, see Appendix \ref{sec:opentransf} for details) can be
solved using a Baxter T-Q equation with an inhomogeneous term, i.e. with the
structure \cite{Cao:2013qxa, Wang2015, Nepomechie:2013ila, 
Belliard:2013aaa, Kitanine:2014swa, Belliard:2015} 
\be
T_{1}\, Q   = \bar{\varphi}\, Q^{[2]} + \varphi\, Q^{[-2]} + \Delta\,,
\label{inhomTQ}
\ee
where $\Delta(u)$ is real analytic (in particular, real for real $u$)
and independent of $Q$.  Indeed, such an 
inhomogeneous term is necessary in order for the function $Q(u)$ to be a {\em 
polynomial} in $u$, i.e. 
$Q(u)=\prod_{j=1}^{N}(u-u_{j})(u+u_{j})$. The analyticity of $T_{1}$ together 
with the T-Q equation (\ref{inhomTQ}) then imply the following Bethe equations for the zeros of 
$Q(u)$
\be
\bar{\varphi}(u_{k})\, Q(u_{k}+i) + \varphi(u_{k})\, Q(u_{k}-i) + 
\Delta(u_{k}) = 0\,.
\label{BEopen}
\ee 
Similarly to the case of the periodic chain, by solving the Bethe equations (\ref{BEopen}), one can obtain $Q$, and therefore 
(through (\ref{inhomTQ})) $T_{1}$.
The transfer matrices
for such models \cite{Sklyanin:1988yz}, constructed using non-diagonal
boundary S-matrices \cite{Ghoshal:1993tm, deVega:1993xi}, still obey
\cite{Mezincescu:1990fc, Mezincescu:1991ke, Zhou:1995zy} the Hirota
equation, albeit in a slightly modified form,
\be
T^{+}_{k}\, T^{-}_{k} - T_{k+1}\, T_{k-1} = T_{2,k}\,, \qquad T_{-1}=0\,,
\qquad k = 0, 1, 2, 
\ldots \,,
\label{Hirotaopen}
\ee
where $T_{2,k}$ is given by the quantum determinant (\ref{qdet}).
This naturally raises the question: does the Hirota equation 
(\ref{Hirotaopen}) admit a Lax
representation with inhomogeneous terms?

We answer this question here in the affirmative. Indeed, we show that 
such a Lax representation is given by (\ref{QIopeninhom})-(\ref{QIIopeninhom})
\be
T_{k+1}\, Q^{[k]} - \bar\varphi^{[k]}\, T_{k}^{-}\, Q^{[k+2]} &=& X_{k}\, 
Q^{[-k-2]}+ \sum_{l=0}^{k}\psi_{l,k}\, \Delta^{[2l-k]}\, T_{l}^{[l-k-1]}
\,, \label{inhomQI}\\
\varphi^{[-k]}\, T_{k-1}\, Q^{[-k-2]} - T_{k}^{-}\, Q^{[-k]} &=& 
-Y_{k}\, Q^{[k]}  - \sum_{l=0}^{k-1}\bar{\psi}_{l,k-1}^{-}\, \Delta^{[k-2l-2]}\, T_{l}^{[k-l-1]}
\,,  \label{inhomQII}
\ee
where $X_{k}, Y_{k}$ and $\psi_{l,k}$ are given by (\ref{XY}) and (\ref{psi}).
The key new point is the appearance of terms containing $\Delta$,
which do not contain $Q$ and therefore are ``inhomogeneous'' terms.
In particular, (\ref{inhomQI}) reduces to (\ref{inhomTQ}) for $k=0$. 
As the equations (\ref{inhomQI}) are still linear, they can be solved
for all $T_{k}$ in terms of $Q$.
We remark that equivalent expressions for
$T_{k}$ in terms of $Q$ were obtained earlier by means of a generating
function \cite{Nepomechie:2013ila}.  An AdS/CFT generalization of this
generating function was proposed in \cite{Zhang:2015fea}, and it was
subsequently used in \cite{Bajnok:2015kfz} to compute wrapping
corrections.

Interestingly, the compatibility of the system 
(\ref{inhomQI})-(\ref{inhomQII}) leads to a family of 
Hirota-like equations (\ref{genHirota})
\be
T_{k+1}\, T_{k-a-1}^{[a]} 
- T_{k}^{-}\, T_{k-a}^{[a+1]}
+ T_{2,k-a}^{[a]}\, T_{a}^{[a-k-1]} = 0\,, \qquad 
a=0\,,1\,, \ldots\,, k-1\,,
\label{newids}
\ee
whose particular case $a=0$ coincides with the original Hirota equation 
(\ref{Hirotaopen}).
To our knowledge, the bilinear relations (\ref{newids}) with $a>0$ are new.  We show 
that these relations are consistent with the Hirota equation (\ref{Hirotaopen})
by first solving the latter to obtain a determinant expression for
$T_{k}$ in terms of $T_{1}$, and by then
judiciously applying Pl\"ucker relations. 

The outline of this paper is as follows.  In Sec.  \ref{sec:periodic}
we briefly review for the periodic spin chain how the compatibility of the
auxiliary linear problem implies the Hirota equation. In Sec. 
\ref{sec:open} we turn to the open spin chain. We present both 
homogeneous and inhomogeneous Lax representations of the Hirota 
equation. We derive the compatibility conditions for the auxiliary problem 
(\ref{inhomQI})-(\ref{inhomQII}), and show that they are satisfied if
the Hirota-like equations (\ref{newids}) are obeyed.  In Sec.
\ref{sec:solvingHirota} we solve the Hirota equation 
(\ref{Hirotaopen}) to obtain a
determinant expression for $T_{k}$ in terms of $T_{1}$, see
(\ref{Tk1}) and (\ref{Tk2}).  In Sec.  \ref{sec:solvingHirotalike} we
use Pl\"ucker relations to show that this solution is also a solution
of the Hirota-like equations.  In Sec.  \ref{sec:discuss} we briefly
discuss our results and we point out some further related problems. 
We briefly review the 
construction of the family of commuting
transfer matrices for integrable periodic and 
open quantum spin chains in appendices \ref{sec:closedtransf} and 
\ref{sec:opentransf}, respectively.

\section{Periodic spin chain}\label{sec:periodic}

It is useful to begin by briefly reviewing how the compatibility of 
the auxiliary linear problem for a periodic spin chain 
(\ref{QI})-(\ref{QII}) with (\ref{QbarQ})
\be
T_{k+1}\, Q^{[k]} - T_{k}^{-}\, Q^{[k+2]} &=& \phi^{[k]}\, 
Q^{[-k-2]} \,, \label{QIagain}\\
T_{k-1}\, Q^{[-k-2]} - T_{k}^{-}\, Q^{[-k]} &=& 
-\bar{\phi}^{[-k]}\, 
Q^{[k]} \,,  \label{QIIagain}
\ee
implies the Hirota equation (\ref{Hirota}).
Multiplying (\ref{QIIagain}) by $T_{k+1}$ gives
\be
T_{k+1}\, T_{k-1}\, Q^{[-k-2]} - T_{k+1}\, T_{k}^{-}\, Q^{[-k]} = 
-\bar{\phi}^{[-k]}\, T_{k+1}\, Q^{[k]} \,.
\label{a}
\ee
On the other hand, performing on (\ref{QIIagain}) the shifts $k \mapsto k+1$ and 
$u\mapsto u+\frac{i}{2}$, and then multiplying the result by $T_{k}^{-}$ gives
\be
T_{k}^{-}\, T_{k}^{+}\, Q^{[-k-2]} - T_{k}^{-}\, T_{k+1}\, Q^{[-k]} = 
-\bar{\phi}^{[-k]}\, T_{k}^{-}\, Q^{[k+2]} \,.
\label{b}
\ee
Subtracting (\ref{a}) from (\ref{b}) yields
\be
\left(T^{+}_{k}\, T^{-}_{k} - T_{k+1}\, T_{k-1} \right)  Q^{[-k-2]}
= \bar{\phi}^{[-k]}\left( T_{k+1}\, Q^{[k]} - T_{k}^{-}\, Q^{[k+2]} 
\right) =   \bar{\phi}^{[-k]}\, \phi^{[k]}\, Q^{[-k-2]}\,,
\label{c}
\ee
where the second equality follows from (\ref{QIagain}).  It is now clear
that (\ref{c}), which expresses the compatibility of (\ref{QIagain}) and
(\ref{QIIagain}) for the function $Q$, implies the Hirota
equation (\ref{Hirota}).  Note also that (\ref{QIIagain}) can be obtained 
from (\ref{QIagain}): performing on (\ref{QIagain}) the shifts $k \mapsto k-1$
and $u\mapsto u+\frac{i}{2}$, we obtain
\be
T_{k}^{+}\, Q^{[k]} -T_{k-1}\, Q^{[k+2]}   =
\phi^{[k]}\, Q^{[-k]} \,,
\label{ItoII}
\ee
which (up to an overall factor $-1$) is the complex conjugate of
(\ref{QIIagain}), assuming that 
$T_{k}(u)^* = T_{k}(u^*)$ is real analytic and $\phi(u)^*=\bar{\phi}(u^*)$. 
In fact, provided that the last two conditions are met, the entire
reasoning can be extended to the general case $Q(u)^*= \bar{Q}(u^*)$.

\section{Open spin chain}\label{sec:open}

For an open spin chain, the corresponding function $\varphi(u)$ 
(\ref{phiopen}) does not 
satisfy the constraint $\bar\phi=\phi^{[-2]}$ (\ref{k0constraint}) 
that follows from  (\ref{Hirota}). Indeed, the Hirota 
equation takes a form slightly different from (\ref{Hirota}), 
namely,
\be
T^{+}_{k}\, T^{-}_{k} - T_{k+1}\, T_{k-1} = T_{2, k}\,, \qquad T_{-1}=0\,,
\qquad k = 0, 1, 2, 
\ldots \,, 
\label{Hirotaopenagain}
\ee
where $T_{2,k}$ is the quantum determinant
\be
T_{2,k} = \prod_{j=0}^{k-1}\varphi^{[k-2j]}\, \bar\varphi^{[2j-k]}\,,
\label{qdet}
\ee 
which satisfies the discrete Laplace equation
\be
T^{+}_{2,k}\, T^{-}_{2,k} =T_{2,k+1}\, T_{2,k-1} \,.
\ee
Since $T_{2,0}=1$, Eq. (\ref{Hirotaopenagain}) with $k=0$ implies that 
\be
T_{0}=1 \,,
\ee
which differs from the first relation of (\ref{k0constraint}).

\subsection{Homogeneous case}

For an open spin chain with {\em diagonal} boundary terms 
$(\xi=0)$, we propose
the following homogeneous auxiliary linear problem
\be
T_{k+1}\, Q^{[k]} - \bar\varphi^{[k]}\, T_{k}^{-}\, Q^{[k+2]} &=& X_{k}\, 
Q^{[-k-2]} \,, \label{QIopenhom}\\
\varphi^{[-k]}\, T_{k-1}\, Q^{[-k-2]} - T_{k}^{-}\, Q^{[-k]} &=& 
-Y_{k}\, 
Q^{[k]} \,,  \label{QIIopenhom}
\ee
where
\be
X_{k} = \prod_{j=0}^{k}\varphi^{[k-2j]}\,, \qquad
Y_{k} = \prod_{j=0}^{k-1}\bar\varphi^{[2j-k]} \,,
\label{XY}
\ee
instead of (\ref{QIagain})-(\ref{QIIagain}). Indeed, following the same 
steps as in the periodic case (\ref{a})-(\ref{c}), we find with the 
help of the simple identities
\be
Y_{k+1}^{+} = \bar\varphi^{[k]}\, Y_{k}\,, \qquad X_{k}\, Y_{k} = 
\varphi^{[-k]}\, T_{2,k}\,,
\label{identities}
\ee
that the compatibility of the linear system 
(\ref{QIopenhom})-(\ref{QIIopenhom}) implies the Hirota equation  
(\ref{Hirotaopenagain}).
Moreover, (\ref{QIIopenhom}) can be obtained from (\ref{QIopenhom}) in the same way
that (\ref{QIIagain}) can be obtained from (\ref{QIagain}), see (\ref{ItoII}).

Eq. (\ref{QIopenhom}) can be solved for $T_{k}$ in 
terms of $Q$. For $k=0$, one readily obtains the usual T-Q equation
\be
T_{1}\, Q   = \bar{\varphi}\, Q^{[2]} + \varphi\, Q^{[-2]}\,,
\ee
as in the periodic case (\ref{BaxTQ}). The result for general values 
of $k$ can alternatively be obtained from a generating function \cite{Nepomechie:2013ila}
\be
{\cal W}_{diag} \equiv (1 - {\cal D} B {\cal D})^{-1}\, (1 - {\cal D} A {\cal D})^{-1} = 
\sum_{k=0}^{\infty} {\cal D}^{k}\, T_{k} \,  {\cal D}^{k} \,,
\label{W1}
\ee
where 
\be
A = \varphi\, \frac{Q^{[-2]}}{Q} \,, \qquad B = \bar{\varphi}\, \frac{Q^{[2]}}{Q} \,,
\label{AB}
\ee
and ${\cal D} =e^{-\frac{i}{2}\partial_{u}}$ implying that ${\cal D} 
f = f^{-} {\cal D}$. In this way, we obtain
\be
T_{k}=\sum_{l=0}^{k}\prod_{j=0}^{k-l-1}B^{[k-1-2j]}\, 
\prod_{i=0}^{l-1}A^{[2l-k-1-2i]}\,.
\label{Tkgendiag}
\ee

\subsection{Inhomogeneous case}

For an open spin chain with {\em non-diagonal} boundary terms 
$(\xi \ne 0)$, we propose
the following inhomogeneous auxiliary linear problem
\be
T_{k+1}\, Q^{[k]} - \bar\varphi^{[k]}\, T_{k}^{-}\, Q^{[k+2]} &=& X_{k}\, 
Q^{[-k-2]}+ \sum_{l=0}^{k}\psi_{l,k}\, \Delta^{[2l-k]}\, T_{l}^{[l-k-1]}
\,, \label{QIopeninhom}\\
\varphi^{[-k]}\, T_{k-1}\, Q^{[-k-2]} - T_{k}^{-}\, Q^{[-k]} &=& 
-Y_{k}\, Q^{[k]}  - \sum_{l=0}^{k-1}\bar{\psi}_{l,k-1}^{-}\, \Delta^{[k-2l-2]}\, T_{l}^{[k-l-1]}
\,,  \label{QIIopeninhom}
\ee
where $X_{k}$ and $Y_{k}$ are given by (\ref{XY}), and $\psi_{l,k}$ 
is given by 
\be
\psi_{l,k} = \prod_{j=0}^{k-l-1}\varphi^{[k-2j]}\,, \qquad
\bar{\psi}_{l,k} = \prod_{j=0}^{k-l-1}\bar{\varphi}^{[2j-k]}\, .
\label{psi}
\ee
For $\Delta=0$, this system evidently reduces to the homogeneous 
system (\ref{QIopenhom})-(\ref{QIIopenhom}). 

Eq.  (\ref{QIopeninhom}) can be used to solve for all $T_{k}$ in terms of $Q$.
The inhomogeneous T-Q equation (\ref{inhomTQ}) is obtained for
$k=0$. The result for general values 
of $k$ can again be alternatively obtained from a generating function \cite{Nepomechie:2013ila}
\be
{\cal W} \equiv \left[1 - {\cal D} (A+B+C) {\cal D} + {\cal D} A {\cal 
D}^{2} B {\cal D} \right]^{-1} = 
\sum_{k=0}^{\infty} {\cal D}^{k}\, T_{k} \,  {\cal D}^{k} \,,
\label{W2}
\ee
where $A$ and $B$ are again given by (\ref{AB}), and $C$ is given by
\be
C=\frac{\Delta}{Q} \,.
\label{C}
\ee 
Note that this generating function reduces to ${\cal W}_{diag}$ 
(\ref{W1}) for $\Delta=0$.

\subsubsection{Compatibility conditions}\label{sec:comp}

We now proceed as in the homogeneous case to derive the compatibility
conditions for the auxiliary linear problem 
(\ref{QIopeninhom})-(\ref{QIIopeninhom})
for the function $Q$.  Multiplying (\ref{QIIopeninhom}) by $T_{k+1}$ gives
\be
T_{k+1}\, T_{k-1}\, \varphi^{[-k]}\, Q^{[-k-2]} - T_{k+1}\, T_{k}^{-}\, 
Q^{[-k]} &=& - Y_{k}\, T_{k+1}\, Q^{[k]}  \non\\ 
&-& \sum_{l=0}^{k-1}\bar{\psi}_{l,k-1}^{-}\, \Delta^{[k-2l-2]}\, T_{k+1}\, T_{l}^{[k-l-1]}
\,.
\label{inhoma}
\ee
On the other hand, performing on (\ref{QIIopeninhom}) the shifts $k \mapsto k+1$ and 
$u\mapsto u+\frac{i}{2}$, and then multiplying the result by $T_{k}^{-}$ gives
\be
T_{k}^{-}\, T_{k}^{+}\, \varphi^{[-k]}\, Q^{[-k-2]} - T_{k}^{-}\, T_{k+1}\, Q^{[-k]} = 
-Y_{k}\, \bar\varphi^{[k]}\, T_{k}^{-}\, Q^{[k+2]} 
- \sum_{l=0}^{k}\bar{\psi}_{l,k}\,\Delta^{[k-2l]}\, T_{k}^{-}\, T_{l}^{[k-l+1]}.
\label{inhomb}
\ee
Subtracting (\ref{inhoma}) from (\ref{inhomb}) yields
\be
\lefteqn{\left(T^{+}_{k}\, T^{-}_{k} - T_{k+1}\, T_{k-1} \right)\varphi^{[-k]}\,  Q^{[-k-2]}
= Y_{k}\left(T_{k+1}\, Q^{[k]} - \bar\varphi^{[k]}\, T_{k}^{-}\, 
Q^{[k+2]}\right)}\\
&&
+ \sum_{l=0}^{k-1}\bar{\psi}_{l,k-1}^{-}\, \Delta^{[k-2l-2]}\, T_{k+1}\, T_{l}^{[k-l-1]}
- \sum_{l=0}^{k}\bar{\psi}_{l,k}\, \Delta^{[k-2l]}\, T_{k}^{-}\, T_{l}^{[k-l+1]} \,. 
\non
\label{inhomc}
\ee
Using (\ref{QIopeninhom}), (\ref{identities}) and (\ref{psi}), we arrive at
\be
\lefteqn{\left(T^{+}_{k}\, T^{-}_{k} - T_{k+1}\, T_{k-1} - T_{2,k}\right)\varphi^{[-k]}\,  Q^{[-k-2]}}\non \\
&&= \sum_{l=0}^{k-1}\bar{\psi}_{l,k-1}^{-}\, \Delta^{[k-2l-2]}\, T_{k+1}\, T_{l}^{[k-l-1]}
- \sum_{l=0}^{k}\bar{\psi}_{l,k}\, \Delta^{[k-2l]}\, T_{k}^{-}\, T_{l}^{[k-l+1]} 
+ \sum_{l=0}^{k}\psi_{l,k}\, Y_{k}\, \Delta^{[2l-k]}\, 
T_{l}^{[l-k-1]} \non\\
&&= \sum_{a=0}^{k-1}\Delta^{[-k+2a]} 
\left(\prod_{j=0}^{a-1}\bar\varphi^{[-k+2j]}\right) H_{k,a}\,, \qquad k 
= 1, 2, \ldots \,,
\label{inhomd}
\ee
where
\be
H_{k,a} = T_{k+1}\, T_{k-a-1}^{[a]} 
- T_{k}^{-}\, T_{k-a}^{[a+1]}
+ T_{2,k-a}^{[a]}\, T_{a}^{[a-k-1]} \,.
\ee 
The compatibility conditions (\ref{inhomd}) are satisfied for  
nonzero $Q$ and $\Delta$ if 
\be
H_{k,a} = 0 \,, \qquad a = 0\,, 1\,, \ldots \,, k-1\,,
\label{genHirota}
\ee
which are precisely the Hirota-like bilinear relations 
(\ref{newids}). (Recall that Eq. (\ref{genHirota}) 
with $a=0$ coincides with the original open-chain Hirota equation  (\ref{Hirotaopen}).)

\subsection{Solving the Hirota equation}\label{sec:solvingHirota}

It is easy to explicitly solve the open-chain Hirota equation
(\ref{Hirotaopenagain}) for $T_{k}$ in terms of $T_{1}$ for small values of
$k$, and to show that the resulting expressions can be conveniently recast in
terms of determinants
\be
T_{2} &=& \left| \begin{array}{cc}
T_{1}^{[1]} & \varphi^{[1]} \\
\bar\varphi^{[-1]} & T_{1}^{[-1]} \end{array} \right| \,, \non\\
T_{3} &=& \left| \begin{array}{ccc}
T_{1}^{[2]} & \varphi^{[2]} & 0\\
\bar\varphi^{[0]} & T_{1}^{[0]} & \varphi^{[0]} \\
0 & \bar{\varphi}^{[-2]} & T_{1}^{[-2]}
\end{array} \right| \,, \non\\
T_{4} &=& \left| \begin{array}{cccc}
T_{1}^{[3]} & \varphi^{[3]} & 0 & 0\\
\bar{\varphi}^{[1]} & T_{1}^{[1]} & \varphi^{[1]} & 0\\
0 & \bar{\varphi}^{[-1]} & T_{1}^{[-1]} & \varphi^{[-1]}\\
0 & 0 & \bar{\varphi}^{[-3]} & T_{1}^{[-3]}
\end{array} \right| \,. 
\ee
This suggests a general determinant expression for $T_{k}$ in 
terms of $T_{1}$ (see also \cite{Krichever:1996qd})
\be
T_{k} = \det (M^{(k)})\,,
\label{Tk1}
\ee
where $M^{(k)}$ is a $k \times k$ matrix whose elements are given by
\be
M^{(k)}_{ij} = T_{1}^{[k+1-2i]} \delta_{ij} + \bar\varphi^{[k+1-2i]}\delta_{i,j+1} + 
\varphi^{[k+1-2i]}\delta_{i,j-1}  \,, \qquad i\,, j = 1\,, \ldots\,, k 
\,.
\label{Tk2}
\ee

We can now verify that (\ref{Tk1}) is the solution of the Hirota 
equation using Jacobi's determinant identity \cite{Krichever:1996qd, 
Hirota:2003}
\be
D[p_{1}, p_{2}| q_{1} q_{2}]\, D = D[p_{1}|q_{1}]\, D[p_{2}|q_{2}] - 
D[p_{1}|q_{2}]\, D[p_{2}|q_{1}]\,,
\label{Jacobi}
\ee
where $D$ is the determinant of a square matrix, and $D[p_{1}, p_{2}, 
\ldots, p_{n}|q_{1}, q_{2}, \ldots, q_{n}]$ denotes the minor determinant obtained 
from the same matrix by removing rows $p_{1}, p_{2}, \ldots, p_{n}$ and columns $q_{1}, q_{2}, 
\ldots, q_{n}$.
Indeed, let us observe that the matrix $M^{(k+1)}$, obtained from 
(\ref{Tk2}), contains $M^{(k-1)}$ as a submatrix
\be
M^{(k+1)} = \left(\begin{array}{c:ccccc:c}
T_{1}^{[k]} & \varphi^{[k]} & 0 & \ldots & 0 & 0 & 0 \\
\hdashline 
\bar{\varphi}^{[k-2]} &  &&&&& 0 \\
\vdots  &&& M^{(k-1)} &&& \vdots \\
0  &&&&&&  \varphi^{[-k+2]} \\
\hdashline  
0 & 0 & 0 & \ldots  & 0 &  \bar{\varphi}^{[-k]} & T_{1}^{[-k]}
\end{array} \right)\,.
\ee
Applying the Jacobi identity (\ref{Jacobi}) to the above $(k+1) 
\times (k+1)$ matrix with 
$p_{1}=q_{1}=1$ and $p_{2}=q_{2}=k+1$, and then using
(\ref{Tk1}), we recover the Hirota equation 
(\ref{Hirota}). (Note that the matrices corresponding to $D[1|k+1]$ 
and $D[k+1|1]$ are either upper or lower triangular, with $\varphi$'s 
along the diagonal.)

We remark that Eq. (\ref{Tk1}) provides an expression for $T_{k}$ in 
terms of $Q$ upon setting $T_{1} = A + B + C$  (see (\ref{AB}),  
(\ref{C}) )
in (\ref{Tk2}). In particular, for the diagonal case $\Delta=0$, the 
result is equivalent to (\ref{Tkgendiag}).

\subsection{Solving the Hirota-like 
equations}\label{sec:solvingHirotalike}

We now demonstrate that the solution (\ref{Tk1}) of the Hirota equation is 
also a solution of the Hirota-like equations (\ref{newids}). The main 
idea is to use Pl\"ucker relations \cite{Krichever:1996qd, 
Hirota:2003}, which are generalizations of Jacobi's identity. 
In this way, we see that the Hirota equation stems from the Jacobi identity, 
while the generalizations of the Hirota equation stem from the 
extension of the Jacobi identity to the Pl\"ucker relations.

Let $X$ be a {\em rectangular} matrix with $n+1$ rows and $r+1$ columns $(n \ge r)$, 
\be
X = \left( \begin{array}{cccc}
X_{0,0} & X_{0,1} & \ldots & X_{0,r}\\
X_{1,0} & X_{1,1} & \ldots & X_{1,r}\\
\vdots & \vdots & \cdots  & \vdots \\
X_{n,0} & X_{n,1} & \ldots & X_{n,r}
\end{array}\right)
\,.
\label{Xmat}
\ee
Following \cite{Krichever:1996qd}, we define  $(i_{0}\,, i_{1}\,, 
\ldots\,, i_{r})$ to be the determinant of the square matrix formed 
by the rows with labels $i_{0}\,, i_{1}\,, \ldots\,, i_{r}$,
\be
\left| \begin{array}{cccc}
X_{i_{0},0} & X_{i_{0},1} & \ldots  & X_{i_{0},r}\\
X_{i_{1},0} & X_{i_{1},1} & \ldots  & X_{i_{1},r}\\
\vdots & \vdots & \cdots  & \vdots \\
X_{i_{r},0} & X_{i_{r},1} & \ldots  & X_{i_{r},r}\\
\end{array} \right| \equiv 
(i_{0}\,, i_{1}\,, \ldots\,, i_{r})\,.
\ee 
It is understood that $(i_{0}\,, i_{1}\,, \ldots\,, i_{r})$ is 
antisymmetric in all the indices, and therefore vanishes if any two 
indices coincide. The Pl\"ucker relations are then given by \cite{Krichever:1996qd}
\be
(i_{0}\,, i_{1}\,, \ldots\,, i_{r})\, (j_{0}\,, j_{1}\,, \ldots\,, 
j_{r}) = \sum_{p=0}^{r} (j_{p}\,, i_{1}\,, \ldots\,, i_{r})\, 
(j_{0}\,,\ldots\,, j_{p-1}\,, i_{0}\,, j_{p+1}\,, \ldots\,, j_{r}) 
\label{Plucker}
\ee
for all $i_{p}, j_{p} \in \{0\,, n \}$ with $p=0, 1, \ldots, r$. 

It is convenient to introduce the notation $[i, p]$ as follows
\be 
(X_{i,0}\,, X_{i,1}\,, \ldots \,, X_{i,r})\Big\vert_{[i, p]} 
= (X_{i,0}\,, X_{i,1}\,, \ldots \,, X_{i,r})\Big\vert_{X_{i,j} = \delta_{j,p}} 
=(\stackrel{\stackrel{0}{\downarrow}}{0}\,, \ldots  0\,, \stackrel{\stackrel{p}{\downarrow}}{1}  \,,  
0\,, \ldots \,, \stackrel{\stackrel{r}{\downarrow}}{0})
\,.
\ee
In other words, $[i, p]$ means 
that the row of matrix $X$ labeled $i$ is given by $(0\,, \ldots  0\,, 1 \,,  0\,, \ldots \,, 0)$, 
where the 1 appears in the column labeled $p$. Similarly, for a 
set of $m$ rows of the matrix $X$, we define
\be
\left( \begin{array}{cccc}
X_{i_{1},0} & X_{i_{1},1} & \ldots  & X_{i_{1},r}\\
X_{i_{2},0} & X_{i_{2},1} & \ldots  & X_{i_{2},r}\\
\vdots & \vdots & \cdots  & \vdots \\
X_{i_{m},0} & X_{i_{m},1} & \ldots  & X_{i_{m},r}\\
\end{array}\right)\left|_{
\left[ \begin{array}{c}
i_{1}, p_{1}\\
i_{2}, p_{2}\\
\vdots\\
i_{m}, p_{m}
\end{array}\right]} \right.
= \left( \begin{array}{cccc}
X_{i_{1},0} & X_{i_{1},1} & \ldots  & X_{i_{1},r}\\
X_{i_{2},0} & X_{i_{2},1} & \ldots  & X_{i_{2},r}\\
\vdots & \vdots & \cdots  & \vdots \\
X_{i_{m},0} & X_{i_{m},1} & \ldots  & X_{i_{m},r}\\
\end{array}\right)\left|_{\begin{array}{c}
X_{i_{1},j}=\delta_{j,p_{1}}\\
X_{i_{2},j}= \delta_{j,p_{2}}\\
\vdots\\
X_{i_{m},j}= \delta_{j,p_{m}} 
\end{array}} \right. \,.
\label{subs}
\ee

We are now ready to show that the solution of the Hirota equation 
(\ref{Hirotaopen}) is also solution to the Hirota-like equations (\ref{newids}). 
We set
\be
r=k\,, \qquad n = r+a+2\,.
\ee 
We then choose the matrix $X$ (\ref{Xmat}) such that its first
$r+1$ rows are given by the matrix $M^{(r+1)}$ in (\ref{Tk2})
\be
\left( \begin{array}{cccc}
X_{0,0} & X_{0,1} & \ldots & X_{0,r}\\
X_{1,0} & X_{1,1} & \ldots & X_{1,r}\\
\vdots & \vdots & \cdots  & \vdots \\
X_{r,0} & X_{r,1} & \ldots & X_{r,r}
\end{array}\right) = M^{(r+1)}
\,,
\label{Xmat2}
\ee
and we choose the remaining rows of $X$ as follows 
\be
\left[ \begin{array}{cc}
r+1, & 0\\
r+2, & r-a\\
r+3, & r-a+1\\
\vdots & \vdots\\
r+a+2,&  r
\end{array}\right]\,,
\label{Xmat3}
\ee
where we have used the notation (\ref{subs}). The Pl\"ucker relations 
(\ref{Plucker}) for this matrix with the following choice of indices \footnote{In \cite{Krichever:1996qd} it is 
assumed that $i_{p}=j_{p}$ for $p\ne 0,1$ in order to reduce the 
number of terms to 3. We do not make this assumption here, but the 
number of terms nevertheless reduces to 3 by virtue of our choice 
(\ref{Xmat3}).}
\be
i_{l} = j_{l} = l \,, \qquad l = 1\,, 2\,, \ldots\,, r-a-1\,, \non \\
j_{0} = 0\,, \quad j_{l} = l\,, \qquad l= r-a\,, r-a+1\,, \ldots\,, r 
\,, \non \\
i_{0} = r+1\,, \quad i_{l}=l+a+2\,, \qquad l= r-a\,, r-a+1\,, 
\ldots\,, r \,,
\label{indices}
\ee
can be shown to coincide (using the identification (\ref{Tk1}))
with the Hirota-like equations 
(\ref{newids}).
We conclude that the solution  (\ref{Tk1}),  (\ref{Tk2}) of the 
original Hirota equation (\ref{Hirotaopen}) also satisfies the 
Hirota-like equations (\ref{newids}), and therefore the latter system 
of equations is consistent with it. We remark that we 
have not succeeded to find a simple transformation that maps the 
Hirota equations (\ref{Hirotaopen}) to the Hirota-like equations (\ref{newids}).

\section{Discussion}\label{sec:discuss}

We have shown that the open-chain $sl(2)$ Hirota equation (\ref{Hirotaopen}) admits a Lax
representation with inhomogeneous terms
(\ref{inhomQI})-(\ref{inhomQII}), thereby demonstrating that the
off-diagonal Bethe ansatz approach \cite{Cao:2013qxa, Wang2015} can be accommodated
within the conventional framework of quantum integrability. 
\footnote{For the special case of diagonal boundary terms 
($\xi=0\,, \Delta=0$), this Lax representation reduces to (\ref{QIopenhom}), 
(\ref{QIIopenhom}), which -- to our knowledge -- is also new.}
In so doing, we have found a family of Hirota-like equations (\ref{newids}) 
which are consistent with the original Hirota equation. 
It is an interesting question whether these Hirota-like equations can 
be obtained directly by fusion.
We expect that (\ref{inhomQI})-(\ref{inhomQII}) is the most general Lax
representation of the Hirota equation with a single function 
$Q(u)$, and therefore it
should describe the most general rank-one integrable quantum system.
It may be interesting to work out the generalization to higher-rank
algebras and superalgebras.  It may also be interesting to investigate
the analogue of such inhomogeneous terms in conformal field theories
and classical integrable systems with boundaries.

\section*{Acknowledgments}

We dedicate this paper to the memory of Petr P. Kulish, who made
seminal contributions to the field of quantum integrability, and who
provided to one of us (RN) a key insight that made possible the
completion of his first foray \cite{Mezincescu:1990fc} into this
field.  DF thanks the partial support of the grants: GAST (INFN),
UniTo-SanPaolo Nr TO-Call3-2012-0088, the ESF Network HoloGrav
(09-RNP-092 (PESC)), the MPNS--COST Action MP1210 and the EU Network
GATIS (no 317089). RN thanks the Bologna INFN theory group for its
warm hospitality,  and Volodya Kazakov for a helpful discussion.  The
work of RN was supported in part by the National Science Foundation
under Grant PHY-1212337, and by a Cooper fellowship.

\appendix

\section{Periodic chain transfer matrices}\label{sec:closedtransf}

We briefly review here the construction of the family of commuting
transfer matrices for a periodic XXX ($sl(2)$-invariant) quantum
spin-1/2 chain with length $N$, whose Hamiltonian is given by
\be
H=\frac{1}{4}\sum_{n=1}^{N}\left(\vec \sigma_{n} \cdot \vec 
\sigma_{n+1} - \id\right)\,, \qquad 
\vec \sigma_{N+1} \equiv \vec \sigma_{1} \,.
\label{Hclosed}
\ee 

The fused $(j, \frac{1}{2})$ $R$-matrices are given by \cite{Kulish:1981gi}
\be
R^{(j,\frac{1}{2})}_{\{a\} \, b}(u) = \chi^{(j)}_{1}(u)\,
P_{\{a\}}^{+}  
\prod_{k=1}^{2j}
R^{(\frac{1}{2},\frac{1}{2})}_{a_{k} b}(u+(k-j-\tfrac{1}{2})i)\, 
P_{\{a\}}^{+} \,, \qquad j=\frac{1}{2}\,, 1\,, \frac{3}{2}\,, \ldots 
\label{fusedRmatrix}
\ee 
where $R^{(\frac{1}{2},\frac{1}{2})}$ is the fundamental 
$sl(2)$-invariant R-matrix (solution of the Yang-Baxter equation)
\be
R^{(\frac{1}{2},\frac{1}{2})} = u + i {\cal P} \,,
\ee
and ${\cal P}$ is the permutation operator.
The $R$-matrices in the product (\ref{fusedRmatrix}) are ordered  
in the order of increasing $k$.
Moreover, 
$P_{\{a\}}^{+}$ is the symmetric projector
\be
P_{\{a\}}^{+} ={1\over (2j)!} 
\prod_{k=1}^{2j}\left(\sum_{l=1}^{k}{\cal P}_{a_{l}, a_{k}} \right) \,,
\label{projector}
\ee
where ${\cal P}_{a_{k}, a_{k}} \equiv 1$. Finally, $\chi^{(j)}_{1}(u)$ is the 
normalization factor
\be
\chi^{(j)}_{1}(u) = \frac{1}{\prod_{k=0}^{2j-2}\left(u+i(j-\frac{1}{2}-k)\right)}\,,
\ee
which removes all trivial zeros. 

The corresponding transfer matrices are given by
\be
t^{(j)}(u) = \tr_{\{a\}} R^{(j,\frac{1}{2})}_{\{a\} \, b^{[N]}}(u) 
\cdots  R^{(j,\frac{1}{2})}_{\{a\} \, b^{[1]}}(u) \,.
\ee
They enjoy the important commutativity property
\be
\left[ t^{(j)}(u)\,, t^{(k)}(v) \right]=0 \,.
\label{commutativity}
\ee
The Hamiltonian (\ref{Hclosed}) is related to the fundamental 
transfer matrix by
\be
H = \frac{i}{2}\frac{d}{du}\ln t^{(\frac{1}{2})}(u)\Big\vert_{u=0} - 
\frac{N}{2}\id\,.
\ee
These transfer matrices are related to the ones discussed in the text
as follows
\be
T_{k}(u) = t^{(\frac{k}{2})}(u-\tfrac{i}{2}) \,, \qquad k = 1\,, 2\,, 
3\,, \ldots 
\label{newT}
\ee
Indeed, these $T_{k}$ obey the Hirota equation (\ref{Hirota}) with 
the functions $\phi$ and $\bar\phi$ given by (\ref{phi}), which indeed satisfy the 
constraint in (\ref{k0constraint}).

\section{Open chain transfer matrices}\label{sec:opentransf}

We now review the construction of the family of commuting transfer
matrices for an open chain.  In addition to R-matrices, we also need
K-matrices (solutions of the boundary Yang-Baxter equation)
\cite{Sklyanin:1988yz, Cherednik:1985vs}. For the XXX case, the general fundamental solution is 
given by \cite{Ghoshal:1993tm, deVega:1993xi}
\be
K^{(\frac{1}{2})}(u; \alpha, \xi_{+}\,, \xi_{-}) = 
 \left(\begin{array}{cc}
                i \alpha+u & u\, \xi_{+}  \\
		 u\,  \xi_{-}     & i \alpha-u
		 \end{array}\right) \,,
\ee 		 
where $\alpha, \xi_{+}\,, \xi_{-}$ are arbitrary boundary parameters. 
The fused $K$-matrices are given by \cite{Mezincescu:1990fc, Mezincescu:1991ke, Zhou:1995zy}
\be
K^{(j)}_{\{a\}}(u) &=& \chi^{(j)}_{2}(u)\, P_{\{a\}}^{+} \prod_{k=1}^{2j} \Bigg\{ \left[ 
\prod_{l=1}^{k-1} R^{(\frac{1}{2},\frac{1}{2})}_{a_{l}a_{k}}
(2u+(k+l-2j-1)i) \right]  \non \\
&& \times K^{(\frac{1}{2})}_{a_{k}}(u+(k-j-\tfrac{1}{2})i) \Bigg\}
P_{\{a\}}^{+} \,, \qquad j=\frac{1}{2}\,, 1\,, \frac{3}{2}\,, \ldots 
\label{fusedKmatrix}
\ee 
where the products of braces $\{ \ldots \}$
are ordered in the order of increasing $k$, and the dependence on the 
boundary parameters has been suppressed. Moreover, 
$\chi^{(j)}_{2}(u)$ is the normalization factor
\be
\chi^{(j)}_{2}(u) = 
\frac{1}{\prod_{k=1}^{4j-3}\left(u+i(j-\frac{k}{2})\right)} \,.
\ee

For simplicity, we consider here the following right and left K-matrices
\be
K^{r (j)}(u)= K^{(j)}(u; \alpha, 0, 0)\,, \qquad
K^{l (j)}(u)= K^{(j)}(-u-i; \beta, \xi, \xi)\,,
\ee 
respectively, with $\alpha, \beta, \xi$ real.
In other words, we choose the right K-matrices to be diagonal, and the left 
K-matrices to be non-diagonal with $ \xi_{+} =  \xi_{-} =  \xi$.

The corresponding open-chain transfer matrices $t^{(j)}(u)$ are given 
by \cite{Sklyanin:1988yz}
\be
t^{(j)}(u) = \tr_{\{a\}} K^{l (j)}_{\{a\}}(u)\,
T^{(j)}_{\{a\}}(u)\, K^{r (j)}_{\{a\}}(u)\,
\hat T^{(j)}_{\{a\}}(u) \,,
\label{transfer}
\ee 
where the monodromy matrices are given by products of $N$ fused $R$-matrices, 
\be
T^{(j)}_{\{a\}}(u) &=& R^{(j,\frac{1}{2})}_{\{a\}\, b^{[N]}}(u) \ldots 
R^{(j,\frac{1}{2})}_{\{a\}\, b^{[1]}}(u) \,, \non \\
\hat T^{(j)}_{\{a\}}(u) &=& R^{(j,\frac{1}{2})}_{\{a\}\, b^{[1]}}(u) \ldots
R^{(j,\frac{1}{2})}_{\{a\}\, b^{[N]}}(u) \,.
\ee 
The corresponding open-chain Hamiltonian, which is obtained 
from the fundamental transfer matrix, is given by
\be
H &=& \frac{i 
(-1)^{N+1}}{2\alpha\beta}\frac{d}{du}t^{(\frac{1}{2})}(u)\Big\vert_{u=0} - N \id \non \\
 &=& \sum_{n=1}^{N-1}\vec \sigma_{n} \cdot \vec \sigma_{n+1} 
 + \frac{1}{\alpha}\sigma^{z}_{1}
 -\frac{1}{\beta}\left(\xi \sigma^{x}_{N} + \sigma^{z}_{N}\right) \,.
\ee 
The transfer matrices (\ref{transfer}) also have the important commutativity 
property (\ref{commutativity}).
We define corresponding $T_{k}$ which are related to (\ref{transfer}) in the same 
way as for the closed chain, namely
\be
T_{k}(u) = t^{(\frac{k}{2})}(u-\tfrac{i}{2}) \,, \qquad k = 1\,, 2\,, 
3\,, \ldots 
\label{newTagain}
\ee
When suitably normalized, these transfer matrices obey 
\cite{Mezincescu:1990fc, Mezincescu:1991ke, Zhou:1995zy}
the open-chain Hirota equation (\ref{Hirotaopenagain}),  (\ref{qdet}) with
\be
\varphi(u) = 
-\frac{1}{u}\left(u+i(\alpha-\tfrac{1}{2})\right)\left(\sqrt{1+\xi^{2}}(u-\tfrac{i}{2})-i\beta\right)(u+\tfrac{i}{2})^{2N+1} \,,
\label{phiopen}
\ee
and $\bar\varphi(u) = \left(\varphi(u^{*})\right)^{*}$.
As noted in Sec. \ref{sec:open}, this function does not satisfy the constraint
(\ref{k0constraint}) for generic values of the boundary parameters. 
When $\xi=0$, both the left and right K-matrices are diagonal, and
the transfer matrix has a $U(1)$ symmetry. If $\xi\ne 0$, then 
this symmetry is broken.

The T-Q equation for the fundamental transfer matrix is given by
(\ref{inhomTQ}) with 
$\varphi(u)$ given by (\ref{phiopen}), and with $\Delta$ given by
\cite{Cao:2013qxa, Wang2015}
\be
\Delta = - 2\left(1-\sqrt{1+\xi^{2}}\right)(u+\tfrac{i}{2})^{2N+1} 
(u-\tfrac{i}{2})^{2N+1}\,,
\ee
which can be derived from basic properties (functional 
relation, crossing symmetry, asymptotic behavior, analyticity) of the transfer 
matrix. Note that $\Delta$ vanishes for $\xi=0$, i.e. when the model has a $U(1)$ 
symmetry. For small values of $N$, one can easily check explicitly  \cite{Nepomechie:2013ila} 
that for each eigenvalue of the transfer matrix, there exists a 
polynomial function $Q(u)=\prod_{j=1}^{N}(u-u_{j})(u+u_{j})$ that satisfies this T-Q equation.


\begin{thebibliography}{10}

\bibitem{Hirota:1981}
R.~Hirota, ``{Discrete Analogue of a Generalized Toda Equation},''
  \href{http://dx.doi.org/10.1143/JPSJ.50.3785}{{\em J. Phys. Soc. Jpn.}
  {\bfseries 50} (1981) 3785--3791}.

\bibitem{Kulish:1981bi}
P.~P. Kulish and E.~K. Sklyanin, ``{Quantum spectral transform method. Recent
  developments},''
{\em Lect. Notes Phys.} {\bfseries 151} (1982) 61--119.

\bibitem{Kirillov:1987zz}
A.~N. Kirillov and N.~Y. Reshetikhin, ``{Exact solution of the integrable XXZ
  Heisenberg model with arbitrary spin. I. The ground state and the excitation
  spectrum},''
\href{http://dx.doi.org/10.1088/0305-4470/20/6/038}{{\em J. Phys.} {\bfseries
  A20} (1987) 1565--1585}.

\bibitem{Bazhanov:1987zu}
V.~V. Bazhanov and N.~{\relax Yu}. Reshetikhin, ``{Critical RSOS Models and
  Conformal Field Theory},''
\href{http://dx.doi.org/10.1142/S0217751X89000042}{{\em Int. J. Mod. Phys.}
  {\bfseries A4} (1989) 115--142}.

\bibitem{Klumper:1992vt}
A.~Klumper and P.~A. Pearce, ``{Conformal weights of RSOS lattice models and
  their fusion hierarchies},''
{\em Physica} {\bfseries A183} (1992) 304.

\bibitem{Kuniba:1993cn}
A.~Kuniba, T.~Nakanishi, and J.~Suzuki, ``{Functional relations in solvable
  lattice models. 1: Functional relations and representation theory},''
  \href{http://dx.doi.org/10.1142/S0217751X94002119}{{\em Int. J. Mod. Phys.}
  {\bfseries A9} (1994) 5215--5266},
\href{http://arxiv.org/abs/hep-th/9309137}{{\ttfamily arXiv:hep-th/9309137
  [hep-th]}}.

\bibitem{Bazhanov:1994ft}
V.~V. Bazhanov, S.~L. Lukyanov, and A.~B. Zamolodchikov, ``{Integrable
  structure of conformal field theory, quantum KdV theory and thermodynamic
  Bethe ansatz},'' \href{http://dx.doi.org/10.1007/BF02101898}{{\em Commun.
  Math. Phys.} {\bfseries 177} (1996) 381--398},
\href{http://arxiv.org/abs/hep-th/9412229}{{\ttfamily arXiv:hep-th/9412229
  [hep-th]}}.

\bibitem{Fioravanti:1995cq}
D.~Fioravanti, F.~Ravanini, and M.~Stanishkov, ``{Generalized KdV and quantum
  inverse scattering description of conformal minimal models},''
  \href{http://dx.doi.org/10.1016/0370-2693(95)01463-2}{{\em Phys. Lett.}
  {\bfseries B367} (1996) 113--120},
\href{http://arxiv.org/abs/hep-th/9510047}{{\ttfamily arXiv:hep-th/9510047
  [hep-th]}}.

\bibitem{Krichever:1996qd}
I.~Krichever, O.~Lipan, P.~Wiegmann, and A.~Zabrodin, ``{Quantum integrable
  systems and elliptic solutions of classical discrete nonlinear equations},''
  \href{http://dx.doi.org/10.1007/s002200050165}{{\em Commun. Math. Phys.}
  {\bfseries 188} (1997) 267--304},
\href{http://arxiv.org/abs/hep-th/9604080}{{\ttfamily arXiv:hep-th/9604080
  [hep-th]}}.

\bibitem{Zabrodin:1998}
A.~V. Zabrodin, ``{Hirota equation and Bethe ansatz},'' {\em Theor. Math.
  Phys.} {\bfseries 116} (1998) 782--819.

\bibitem{Gromov:2008gj}
N.~Gromov, V.~Kazakov, and P.~Vieira, ``{Finite Volume Spectrum of 2D Field
  Theories from Hirota Dynamics},''
  \href{http://dx.doi.org/10.1088/1126-6708/2009/12/060}{{\em JHEP} {\bfseries
  12} (2009) 060},
\href{http://arxiv.org/abs/0812.5091}{{\ttfamily arXiv:0812.5091 [hep-th]}}.

\bibitem{Gromov:2009tv}
N.~Gromov, V.~Kazakov, and P.~Vieira, ``{Exact Spectrum of Anomalous Dimensions
  of Planar N=4 Supersymmetric Yang-Mills Theory},''
  \href{http://dx.doi.org/10.1103/PhysRevLett.103.131601}{{\em Phys.Rev.Lett.}
  {\bfseries 103} (2009) 131601},
\href{http://arxiv.org/abs/0901.3753}{{\ttfamily arXiv:0901.3753 [hep-th]}}.

\bibitem{Kulish:1981gi}
P.~P. Kulish, N.~{\relax Yu}. Reshetikhin, and E.~K. Sklyanin, ``{Yang-Baxter
  Equation and Representation Theory. 1.},''
\href{http://dx.doi.org/10.1007/BF02285311}{{\em Lett. Math. Phys.} {\bfseries
  5} (1981) 393--403}.

\bibitem{Baxter1982}
R.~J. Baxter, {\em Exactly Solved Models in Statistical Mechanics}.
\newblock Academic Press, 1982.

\bibitem{Cao:2013qxa}
J.~Cao, W.-L. Yang, K.~Shi, and Y.~Wang, ``{Off-diagonal Bethe ansatz solution
  of the XXX spin-chain with arbitrary boundary conditions},''
  \href{http://dx.doi.org/10.1016/j.nuclphysb.2013.06.022}{{\em Nucl. Phys.}
  {\bfseries B875} (2013) 152--165},
\href{http://arxiv.org/abs/1306.1742}{{\ttfamily arXiv:1306.1742 [math-ph]}}.

\bibitem{Wang2015}
Y.~Wang, W.-L. Yang, J.~Cao, and K.~Shi, {\em Off-Diagonal Bethe Ansatz for
  Exactly Solvable Models}.
\newblock Springer, 2015.

\bibitem{Nepomechie:2013ila}
R.~I. Nepomechie, ``{An inhomogeneous T-Q equation for the open XXX chain with
  general boundary terms: completeness and arbitrary spin},''
  \href{http://dx.doi.org/10.1088/1751-8113/46/44/442002}{{\em J.Phys.}
  {\bfseries A46} (2013) 442002},
\href{http://arxiv.org/abs/1307.5049}{{\ttfamily arXiv:1307.5049 [math-ph]}}.

\bibitem{Belliard:2013aaa}
S.~Belliard and N.~Crampe, ``{Heisenberg XXX Model with General Boundaries:
  Eigenvectors from Algebraic Bethe Ansatz},''
  \href{http://dx.doi.org/10.3842/SIGMA.2013.072}{{\em SIGMA} {\bfseries 9}
  (2013) 072},
\href{http://arxiv.org/abs/1309.6165}{{\ttfamily arXiv:1309.6165 [math-ph]}}.

\bibitem{Kitanine:2014swa}
N.~Kitanine, J.~M. Maillet, and G.~Niccoli, ``{Open spin chains with generic
  integrable boundaries: Baxter equation and Bethe ansatz completeness from
  separation of variables},''
  \href{http://dx.doi.org/10.1088/1742-5468/2014/05/P05015}{{\em J. Stat.
  Mech.} {\bfseries 1405} (2014) P05015},
\href{http://arxiv.org/abs/1401.4901}{{\ttfamily arXiv:1401.4901 [math-ph]}}.

\bibitem{Belliard:2015}
S.~Belliard and R.~A. Pimenta, ``{Slavnov and Gaudin-Korepin formulas for
  models without U(1) symmetry: the XXX chain on the segment},'' {\em J. Phys.}
  {\bfseries A49} (2016) 17LT01,
  \href{http://arxiv.org/abs/1507.03242}{{\ttfamily arXiv:1507.03242
  [math-ph]}}.

\bibitem{Sklyanin:1988yz}
E.~Sklyanin, ``{Boundary Conditions for Integrable Quantum Systems},''
\href{http://dx.doi.org/10.1088/0305-4470/21/10/015}{{\em J.Phys.} {\bfseries
  A21} (1988) 2375--289}.

\bibitem{Ghoshal:1993tm}
S.~Ghoshal and A.~B. Zamolodchikov, ``{Boundary S matrix and boundary state in
  two-dimensional integrable quantum field theory},''
  \href{http://dx.doi.org/10.1142/S0217751X94001552,
  10.1142/S0217751X94001552}{{\em Int.J.Mod.Phys.} {\bfseries A9} (1994)
  3841--3886}, \href{http://arxiv.org/abs/hep-th/9306002}{{\ttfamily
  arXiv:hep-th/9306002 [hep-th]}}.

\bibitem{deVega:1993xi}
H.~de~Vega and A.~Gonzalez-Ruiz, ``{Boundary K matrices for the XYZ, XXZ and
  XXX spin chains},'' {\em J.Phys.} {\bfseries A27} (1994) 6129--6138,
\href{http://arxiv.org/abs/hep-th/9306089}{{\ttfamily arXiv:hep-th/9306089
  [hep-th]}}.

\bibitem{Mezincescu:1990fc}
L.~Mezincescu, R.~I. Nepomechie, and V.~Rittenberg, ``{Bethe ansatz solution of
  the Fateev-Zamolodchikov quantum spin chain with boundary terms},''
\href{http://dx.doi.org/10.1016/0375-9601(90)90016-H}{{\em Phys. Lett.}
  {\bfseries A147} (1990) 70--78}.

\bibitem{Mezincescu:1991ke}
L.~Mezincescu and R.~I. Nepomechie, ``{Fusion procedure for open chains},''
{\em J. Phys.} {\bfseries A25} (1992) 2533--2544.

\bibitem{Zhou:1995zy}
Y.-K. Zhou, ``{Row transfer matrix functional relations for Baxter's eight
  vertex and six vertex models with open boundaries via more general reflection
  matrices},'' \href{http://dx.doi.org/10.1016/0550-3213(95)00553-6}{{\em Nucl.
  Phys.} {\bfseries B458} (1996) 504--532},
\href{http://arxiv.org/abs/hep-th/9510095}{{\ttfamily arXiv:hep-th/9510095
  [hep-th]}}.

\bibitem{Zhang:2015fea}
X.~Zhang, J.~Cao, S.~Cui, R.~I. Nepomechie, W.-L. Yang, K.~Shi, and Y.~Wang,
  ``{Bethe ansatz for an AdS/CFT open spin chain with non-diagonal
  boundaries},'' \href{http://dx.doi.org/10.1007/JHEP10(2015)133}{{\em JHEP}
  {\bfseries 10} (2015) 133},
\href{http://arxiv.org/abs/1507.08866}{{\ttfamily arXiv:1507.08866 [hep-th]}}.

\bibitem{Bajnok:2015kfz}
Z.~Bajnok and R.~I. Nepomechie, ``{Wrapping corrections for non-diagonal
  boundaries in AdS/CFT},''
  \href{http://dx.doi.org/10.1007/JHEP02(2016)024}{{\em JHEP} {\bfseries 02}
  (2016) 024},
\href{http://arxiv.org/abs/1512.01296}{{\ttfamily arXiv:1512.01296 [hep-th]}}.

\bibitem{Hirota:2003}
R.~Hirota, ``{Determinants and Pfaffians: How to obtain N-soliton solutions
  from 2-soliton solutions},'' {\em RIMS} {\bfseries 1302} (2003) 220--242.

\bibitem{Cherednik:1985vs}
I.~V. Cherednik, ``{Factorizing Particles on a Half Line and Root Systems},''
  \href{http://dx.doi.org/10.1007/BF01038545}{{\em Theor. Math. Phys.}
  {\bfseries 61} (1984) 977--983}.
[Teor. Mat. Fiz.61,35(1984)].

\end{thebibliography}

\providecommand{\href}[2]{#2}\begingroup\raggedright\endgroup

\end{document}